\documentclass[useAMS,usenatbib]{mn2e}
\usepackage{latexsym}
\usepackage{lscape}
\usepackage{rotating}
\usepackage{amsmath}
\usepackage[T1]{fontenc}
\usepackage{aecompl}

\def\msun{\hbox{M$_\odot$}}

\def\cm3{\hbox{cm$^{-3}$}}

\input psfig.sty
\input epsf.sty
\usepackage{graphicx}
\usepackage{epsfig}
\title[Stellar Rotational effects on cluster CMDs]
{Apparent Age Spreads in Clusters and the Role of Stellar Rotation}
\author[Niederhofer et al.] {F. Niederhofer$^{1,2}$\thanks{FN: niederhofer$@$usm.lmu.de}, C. Georgy$^3$, N. Bastian$^{4}$, and S. Ekstr\"om$^5$\\
$^{1}$ Excellence Cluster Origin and Structure of the Universe, Boltzmannstr. 2, D-85748 Garching bei M\"unchen, Germany \\
$^{2}$ Universit\"ats-Sternwarte M\"unchen, Scheinerstra{\ss}e 1, D-81679 M\"unchen, Germany \\
$^{3}$ Astrophysics group, EPSAM, Keele University, Lennard-Jones Labs, Keele, ST5 5BG, UK \\
$^{4}$ Astrophysics Research Institute, Liverpool John Moores University, 146 Brownlow Hill, Liverpool L3 5RF, UK \\
$^{5}$ Geneva Observatory, University of Geneva, Maillettes 51, 1290, Sauverny, Switzerland 
}
\date{Accepted. Received; in original form}
\pagerange{\pageref{firstpage}--\pageref{lastpage}}
\pubyear{2015}
\begin{document}
\maketitle
\label{firstpage}
\begin{abstract}

We use the Geneva \textsc{Syclist} isochrone models that include the effects of stellar rotation to investigate the role that rotation has on the resulting colour-magnitude diagram (CMD) of young and intermediate age clusters.  We find that if a distribution of rotation velocities exists within the clusters, rotating stars will remain on the main sequence (MS) for longer, appearing to be younger than non-rotating stars within the same cluster.  This results in an extended main sequence turn-off (eMSTO) that appears at young ages ($\sim30$~Myr) and lasts beyond 1~Gyr.  If this eMSTO is interpreted as an age spread, the resulting age spread is proportional to the age of the cluster, i.e. young clusters ($<100$~Myr) appear to have small age spreads (10s of Myr) whereas older clusters ($\sim1$~Gyr) appear to have much large spreads, up to a few hundred Myr.  We compare the predicted spreads for a sample of rotation rates to observations of young and intermediate age clusters, and find a strong correlation between the measured 'age spread' and the age of the cluster, in good agreement with models of stellar rotation.  This suggests that the 'age spreads' reported in the literature may simply be the result of a distribution of stellar rotation velocities within clusters.

\end{abstract}
\begin{keywords} galaxies: star clusters: general -- galaxies: individual: LMC -- Hertzsprung--Russell and colour--magnitude diagrams -- stars: rotation

\end{keywords}

\section{Introduction}

The phenomenon of extended main sequence turn-offs (eMSTOs), where the MSTO in colour-magnitude diagrams is broader than predicted by a single isochrone (including the effects of binarity and photometric errors), is a common feature of intermediate age ($1-2$~Gyr; e.g. \citealt{MackeyBrobyNielsen07}) and apparently even younger ($\sim300$~Myr; \citealt{Milone15};\citealt{Correnti15}) massive stellar clusters. The origin of this feature is still highly debated in the literature (e.g. \citealt{Li14,Goudfrooij15}). 

One interpretation, which disagrees with the long standing paradigm of star clusters being simple stellar populations (SSP), is that there is an age spread inside the clusters of the order of $\sim200-700$~Myr (e.g. \citealt{Goudfrooij09, Goudfrooij11a, Goudfrooij11b, Goudfrooij14, Goudfrooij15, Rubele13, Correnti14}).  As a scenario to form such clusters, \citet{Goudfrooij11b, Goudfrooij14} suggested that a massive cluster initially emerges as an SSP from a near-instantaneous star forming burst, which is then followed by a Gaussian-like extended star formation episode. The second generation stars are formed out of a combination of the ejecta of evolved first generation stars and large amounts of accreted gas from the surroundings. 
The authors then invoke strong cluster dissolution, due to tidal forces and cluster expansion, which preferentially removes 1st generation stars. By an age of $1-2$~Gyr, the clusters have removed the majority of their first generation stars, leaving nearly only the 2nd generation with an extended star-formation history (SFH). While potentially plausible for the intermediate age clusters, this scenario does not appear to be consistent with  the inferred SFH of the younger cluster NGC~1856 ($\sim300$~Myr; \citealt{Milone15}), where the role of cluster dissolution is expected to be very small (e.g., \citealt{Baumgardt13}) and two distinct peaks in the SFH are not seen. 

Several studies have aimed to search for signs for prolonged star formation histories or ongoing star formation in young massive clusters (e.g. \citealt{Bastian13, Cabrera-Ziri14, Niederhofer15a}, Cabrera-Ziri et al. in prep.), however no such extended or ongoing bursts have yet been found. Theses non-detections call into question the interpretation of age spreads as the cause of extended MSTOs.

Apart from age spreads, other scenarios have been put forward. \citet{BastianDeMink09} were the first to suggest that the eMSTO feature may be due to stellar rotation. In their paper they computed the effect that rotation has on the surface colour and luminosity of stars that rotate at various velocities and were able to reproduce a broadening in the turn-off similar to the observed ones. However, \citet{BastianDeMink09} did not include the effect that rotation has on the interior of the stars, e.g. a longer MS life time due to internal mixing. \citet{Girardi11} criticised the proposed rotation scenario and showed that the prolonged lifetime of moderately rotating stars can compensate for the shift in colour and magnitude, leading to a tight MSTO. In contrast, \citet{Yang13} found in their study that the extended MSTO feature can, at least partially, be explained by stellar rotation, based on calculations performed with the Yale Rotational Stellar Evolutionary Code (YREC, \citealt{Pinsonneault89,YangBi07}).

More recently, studies have investigated the effects of rotation on the MSTO (and colour-magnitude diagrams more generally) based on the Geneva rotating stellar models \citep{Georgy13}.  \citet{BrandtHuang15a} studied the CMD of the Praesepe and Hyades, two nearby Galactic open clusters ($\sim$800 Myr) that both have turn-offs which are not in agreement with a coeval stellar population. They showed that the spread in their turn-offs can be explained with rotating stars. The authors have expanded their models to older isochrones, up to 2.5 Gyrs \citep{BrandtHuang15b}. The models predict an extended MSTO whose width is a function of the cluster age with a maximum spread at around $1-1.5$~Gyr. A comparison of their models with four extended MSTO clusters show that stars with the same age but different rotation rates and inclinations can reproduce the spread in the observed turn-off. 

Finally, \citet{Niederhofer15b} have found a strong correlation between the width of the 'inferred age spread' and the age of the cluster, with younger clusters showing smaller age spreads than older clusters below $\sim1.5$~Gyr and the trend inverting beyond this age.  The correlation coefficient between the age and FWHM is much stronger than between the FWHM and present day mass or escape velocity (or even the 'corrected' values from \citet{Goudfrooij14}) that have been used to argue for a causal connection between the cluster properties and age spreads. It is interesting to note that the younger cluster NGC~1856 continues this trend to lower ages, having a much smaller inferred spread ($\Delta t  = 100-130$~Myr) than its older intermediate age clusters ($\Delta t=300-700$~Myr).

Stars are known to rotate since Galileo Galilei and Johannes Fabricius observed the Sun's spots around 1610. Very early after the understanding of the stellar constitution by \citet{Eddington1916a,Eddington1917a,Eddington1918a}, the need to adapt the internal structure equations to the case of stellar rotation was studied \citep{Milne1923a,vonZeipel1924a,Eddington1925a}. The establishment of a large scale circulation in the interior of rotating stars has been presented first by \citet{Eddington1929a}, and formalised later by \citet{Sweet1950a}. This circulation ({\sl Eddington-Sweet} or {\sl meridional circulation}), is responsible for a mixing inside the star that adds to the mixing by convective movements. The effects of rotation have been included in stellar evolution codes since the 1970's \citep{Kippenhahn1970a,Endal1976a}. After the observations of enrichments in N \citep{Lyubimkov1984a} and He \citep{Herrero1992a} in OB stars, with larger enrichments in more massive stars, rotation was rapidly identified as the cause and started to be seen as a necessary ingredient in stellar models. While the numerical details of its implementation in the various stellar evolution codes differ in many ways, the general effects are always the same: 1) an internal mixing that modifies the chemical composition at the surface and increases the MS lifetime, 2) hydrostatic modifications on the structure. Both effects have strong consequences on the stellar evolutionary paths.

\section{\textsc{Syclist} Models}
\label{sec:models}

\begin{figure}
\includegraphics[width=8.5cm]{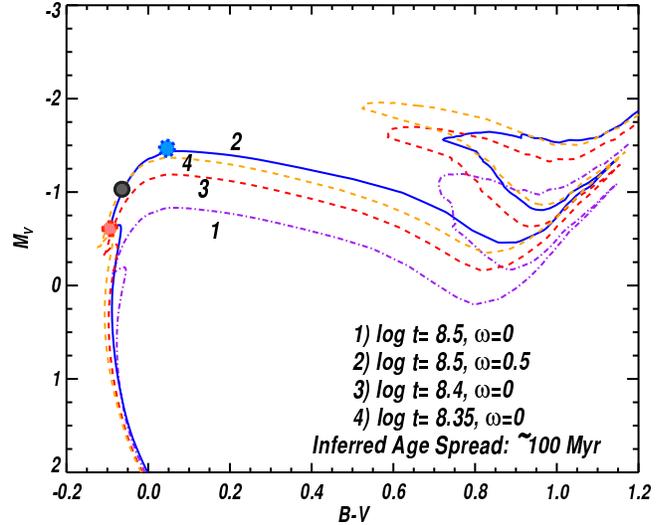}
\caption{Four isochrones from the \textsc{Syclist} models.  The isochrone of a rotating star ($\omega=0.5$) of age log t=8.5 closely resembles that of a non-rotating younger isochrone of an age log t=8.35 at the main-sequence turn off.  Hence, a distribution of rotation rates within a young/intermediate age cluster has a similar effect as an age spread on the MSTO portion of the CMD. The points mark specific points in the evolution of the star and are described in the text.}
\label{fig:isochrones}
\end{figure}

In this study, we used the stellar population synthesis code \textsc{Syclist} \citep{Georgy14}, designed to be used with recent grids of models computed with the Geneva stellar evolution code \citep{Ekstrom12,Georgy13,Georgy13b}. For the present work, the grids presented in \citet{Georgy13} are used and we adopt a metallicity of Z=0.006. They cover a mass range between $1.7$ and $15\,\text{M}_\odot$ and initial rotation rates between $0$ and $95\%$ of the critical one\footnote{We use here the notation $\omega = \Omega / \Omega_\text{crit}$ for the ratio of the actual surface angular velocity to the critical one. The critical angular velocity accounts for the effect of the deformation of the shape of the stellar surface produced by rotation (cf. \citealt{Maeder09}):
\begin{equation*}
\Omega_\text{crit} = \sqrt{\frac{8GM}{27R_\text{p, crit}^3}}
\end{equation*}
where $R_\text{p, crit}$ is the polar radius at the critical velocity.}. 
They are thus well designed for young and intermediate age clusters between $\sim 20\,\text{Myr}$ and $\sim 1\,\text{Gyr}$, and allow one to consider an initial velocity distribution.
Before the final implementation of the grids in \citet{Ekstrom12} and \citet{Georgy13} was done, several tests with various prescriptions for rotation, including angular momentum transport and rotational mixing, were performed. The prescription which reproduces best observational features was finally selected.

\textsc{Syclist} is built to compute interpolated models, isochrones, and synthetic clusters. By summing up clusters of various ages (with different weights), it is also able to compute stellar populations with different SFHs. The exact way stellar populations are computed is described in \citet{Meynet15}. In \textsc{Syclist}, the colours and magnitudes are obtained from the luminosity, surface gravity, effective temperature and metallicity following the work of \citet{Worthey2011}. In this paper, we use the observationally determined velocity distribution of \citet{Huang10}.

The effect of rotation on CMDs of clusters was already studied in previous works. Due to the different approaches followed so far, we summarise here briefly the main differences between these studies and the present one. In \citet{BastianDeMink09}, the effects of rotation where parametrised from a test case, and then applied to non-rotating isochrones to mimic the effect of rotation. The main problem of this approach is that the modification of the lifetime of rotating stars is not accounted for (rotating stars have a longer lifetime than non-rotating ones, at least in the framework of the grids by \citealt{Georgy13}). In \citet{Girardi11}, rotation was accounted for the same way than we do here. However, differences can be noted about the choice of the diffusion coefficients used, as well as the convective overshooting. Moreover, they focused on clusters with a higher age than here. In \citet{Yang13}, the transport of angular momentum inside the star is treated as a purely diffusive process, contrarily to our approach where the advective term in the equation of angular momentum transport \citep[see][]{Zahn92} is accounted for.

\subsection{Determining the Expected 'Age Spread'}
\label{sec:mod_delage}

\begin{figure}
\includegraphics[width=8.5cm]{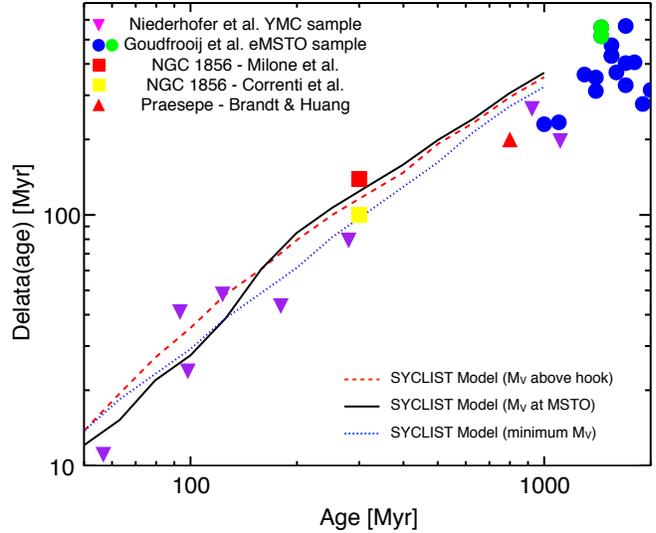}
\caption{The predicted and observed relation between the inferred age spread in young and intermediate age clusters and their age. The dotted (blue), the dashed (red) and the solid (black) lines show the prediction of the \textsc{Syclist} models if rotation-induced main sequence turn-off spreads are inferred to be age spreads based on the differences in magnitudes at three different points along the isochrone (see text).  A key prediction of the rotational models is that the inferred age spread should increase with cluster age (until magnetic breaking becomes important). The filled points show observations of young and intermediate age clusters where age spreads have been inferred based on the MSTOs. The data are well reproduced by the predictions of rotation.}
\label{fig:results}
\end{figure}

\begin{figure}
\includegraphics[width=8.5cm]{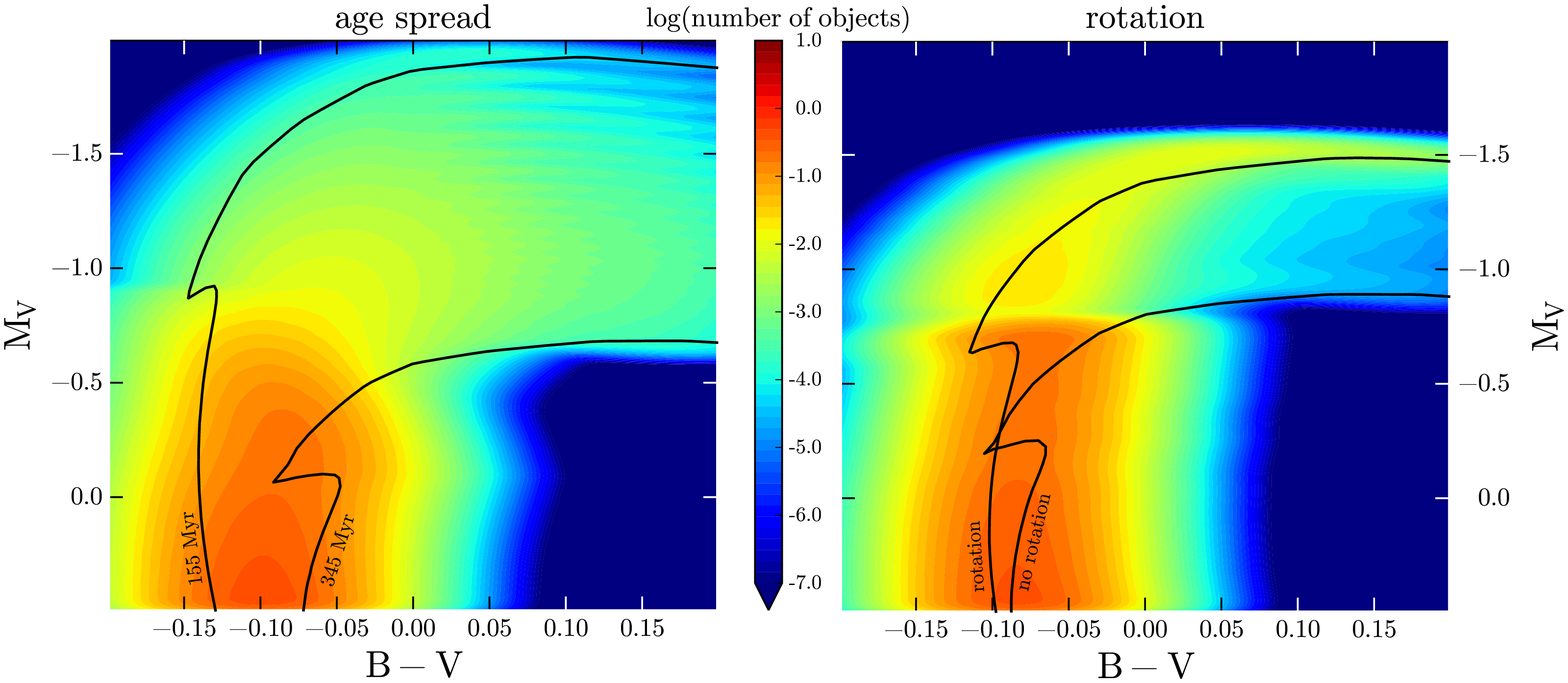}
\includegraphics[width=8.5cm]{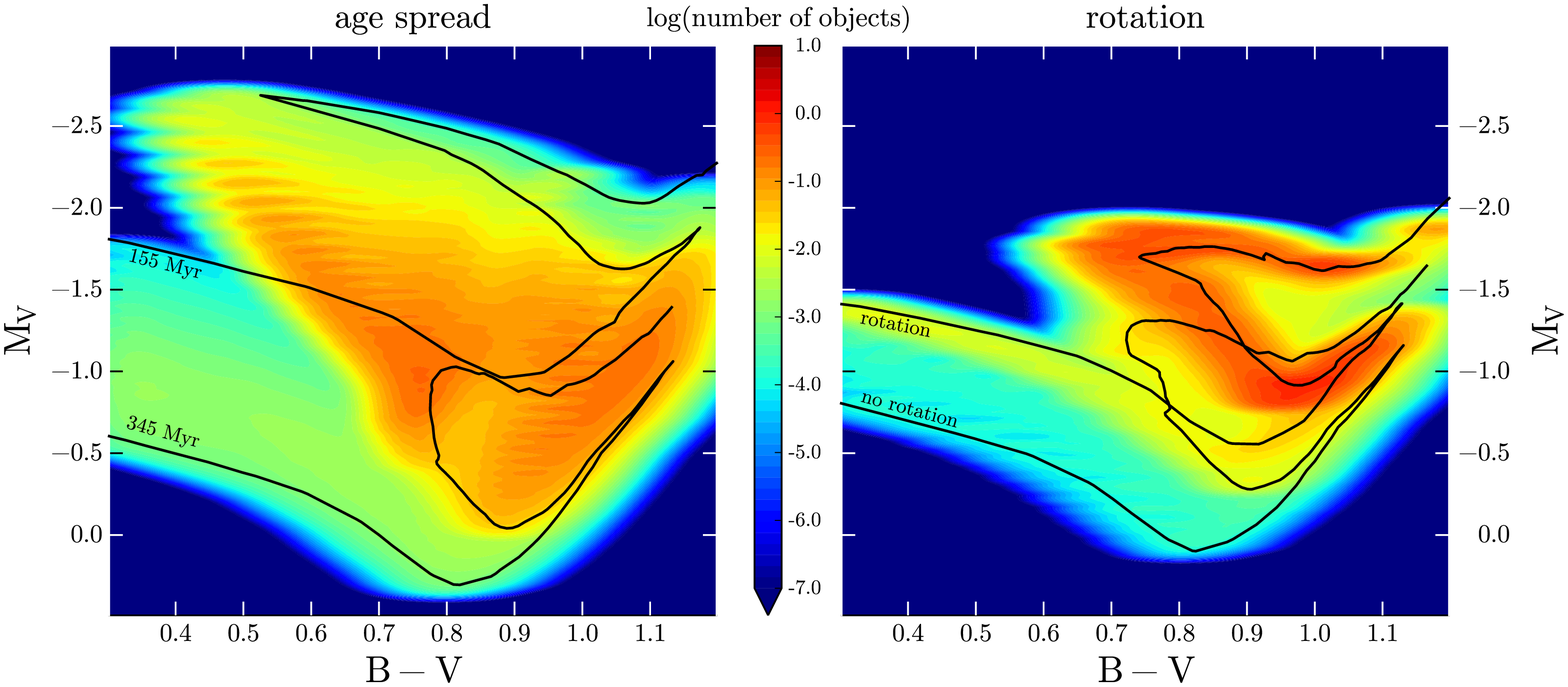}
\caption{{\bf Upper panels:} Hess diagrams centred on the MSTO features of two artificial clusters.
\textit{Left panel:} A cluster constructed from stars with different ages. The ages are from a Gaussian distribution with a peak age of 250 Myr and a standard deviation of 50 Myr and are sampled at 11 equally spaced times. Superimposed as black lines are two isochrones at 155 and 345 Myr.
\textit{Right panel:} A coeval cluster but with stars rotating at various velocities. The age of this population is 300 Myr. The rotation velocity distribution is taken from \citet{Huang10}. The black lines are isochrones for non-rotating stars ($\omega=0.0$) and stars that rotate with half of their break up velocity ($\omega=0.5$). 
Both diagrams are blurred with a Gaussian with a dispersion of 0.03 in colour and 0.02 in magnitude to account for photometric noise. Note that the colour scale in both panels is in logarithmic units.  {\bf Lower panels:} The same as the upper panels but now centred on the post-MS tracks of the cluster CMD.}
\label{fig:hess-msto}
\end{figure}

As shown in \citet[][and references therein]{Georgy14} rotation at young to intermediate ages causes heightened mixing within stars, allowing more hydrogen to enter the core, i.e. rotating stars stay on the MS for longer than non-rotating stars (see also \citealt{Girardi11}). This manifests as an eMSTO that can mimic an age spread within clusters.  

In order to estimate what effect a rotation-induced MSTO spread would be if it was inferred to be an age spread, we performed the following tests, shown in Fig.~\ref{fig:isochrones}. We first adopt an isochrone of a given age, e.g., log t = 8.5, and $\omega=0.0$ (i.e., a non-rotating isochrone), and compare it to an isochrone of the same age but for rotating stars with $\omega=0.5$ (which are labelled isochrones 1 and 2 in Fig.~\ref{fig:isochrones}, respectively). We then over-plotted non-rotating isochrones for younger ages (labelled as isochrones 3 and 4 in Fig.~\ref{fig:isochrones}) until the youngest isochrone that best reproduced the log t=8.5, $\omega=0.5$ isochrone was found.  As shown in Fig.~\ref{fig:isochrones} in the MSTO region of the CMD, rotation manifests itself in a way that is nearly identical to that of an age spread.

We carried out the above test in an automated way for a range of ages, from 50~Myr to 1~Gyr, and derived the inferred spread if the rotation induced broadening was interpreted as an age spread. For this we selected a specific point along the non-rotating and rotating isochrones and searched for the closest match in magnitude. In order to be more objective, we performed the analysis three times using three different points along the isochrones. The locations of these points are indicated as circles along isochrone 2 in Fig.~\ref{fig:isochrones}. 
The first one is the brightest point (minimum $M_V$) after the stars have left the MS (blue dotted circle). 
As the second position in the CMD we selected the evolutionary MSTO (red dashed circle). As the last position we chose the point where the isochrone passes right above the local brightness maximum of the hook-like feature just before the MSTO (solid black circle).

The results are shown as a dotted (blue), dashed (red) and solid (black) line in Fig.~\ref{fig:results}. 
All three choices of the specific points along the isochrones yield similar results.
The inferred age spread is proportional to the age of the cluster, with younger clusters showing smaller age spreads than older clusters.  For clusters $\leq100$~Myr, age spreads of a few 10s of Myr would be expected, whereas for intermediate age clusters ($\sim1$~Gyr), age spreads in excess of 300~Myr are expected. Note the similar relation of the observations. 

For older ages, rotation-induced spreads are expected to decrease at some point. The stars at the turn-off have sufficiently low masses to posses convective envelopes generating magnetized winds. Through these winds the stars loose angular momentum which slows down the stars. This happens at a turn-off mass of $\sim$1.3 M$_{\odot}$ and is referred to as the break in the "Kraft curve" \citep{Kraft67}.

Based on the results of \citet{Niederhofer15b} this appears to happen around $\sim1.5$~Gyr, although \citet{BrandtHuang15b} suggest that this will happen somewhat earlier, at an age of $\sim1-1.5$~Gyr.

We would like to mention here that we used for our analysis rotating isochrones with an initial rotation rate $\omega$=0.5 which is not an extreme value compared to what is measured for B stars (e.g. \citealt{Huang10}). The results that we obtained are valid as long as the initial rotation distribution
roughly covers the interval from $\omega$=0 to $\omega$=0.5. If the initial rotation velocities for specific reasons is peaked towards a single value, the eMSTO features cannot be explained by stellar rotation.

Besides stellar rotation, also convective overshooting modifies the isochrones and could affect our results. However, changing the overshooting in the models would only shift both isochrones, rotating and non-rotating, towards higher luminosities in the CMD, and therefore the effect on our results is minor. As the masses and compositions of the stars at the MSTO are nearly identical, variations of overshooting along the MSTO stars is not expected. 

\subsection{Effect on the Post-Main Sequence Features}
\label{sec:post-ms}

An age spread within a cluster and a range of stellar rotation velocities will not only effect the MSTO but also the post-MS features in the CMD. To test how these two scenarios are reflected in the structure of a CMD, we created two synthetic clusters using the \textsc{Syclist} models and compared their relative stellar densities in colour-magnitude space(Hess diagram). The first cluster is constructed out of non-rotating stars with a Gaussian-like age distribution that has a peak age at 250 Myr and a standard deviation of 50 Myr. The second cluster consists of a coeval stellar population with an age of 300 Myr, however, with stars rotating at various velocities. The rotation distribution was taken from the observationally determined distribution of \citet{Huang10}. The upper panels in Fig.~\ref{fig:hess-msto} show the $B-V$ vs $M_V$ Hess diagram of the two artificial clusters centred on the MSTO. The left panel shows the age spread cluster, whereas the cluster with the rotating stars is displayed in the right panel. The structure is slightly different in both models, however, the location of the turn-off itself (denoted by the top of the high density region in red) is nearly at the same magnitude and colour in both cases. Hence, age spreads and a rotation distribution can lead to very similar MSTO features.

The post-MS morphologies, however, are very different in the two scenarios, as shown in the bottom panels of Fig.~\ref{fig:hess-msto}. In the age spread cluster, the post-MS tracks extend over a large magnitude range (Fig. \ref{fig:hess-msto} left) whereas the evolved stars in the cluster with coeval rotating stars show a more compact morphology in colour-magnitude space (Fig. \ref{fig:hess-msto} right). 

The fact that the morphology of the post-MS differs in the age spread and rotation model can help to discriminate between the two scenarios. We note that \citet{BastianSilva13} focussed their study largely on the post-MS features of NGC~1866 and 1856, and found relatively small age spreads, i.e. $\sigma < 35$~Myr for both clusters.  On the other hand, \citet{Correnti15} and \citet{Milone15} focussed on the MSTO of NGC~1856 and found evidence for a much more extended SFH, $90-130$~Myr.  This apparent contradiction is predicted if rotation is the underlying cause of the observed extended MSTOs.

\section{Comparison with Measurements from the Literature}
\label{sec:comparison}

A number of papers have reported age spreads based on eMSTOs in young and intermediate age clusters (e.g., \citealt{Milone09, Milone15, Goudfrooij14, Correnti15}).  In Fig.~\ref{fig:results} we over-plot the data from the literature on the expectations from the analysis discussed in \S~\ref{sec:mod_delage}. The filled circles are intermediate age clusters from the sample of \citet{Goudfrooij14}, where we have adopted their age and FWHM of the age distribution for each cluster in the LMC (blue) and SMC (green). The filled (red) square represents NGC~1856 from the recent study of \citet{Milone15} and the (yellow) filled square represents the same cluster but the age spread is from \citet{Correnti15}\footnote{\citet{Correnti15} prefer a two-component single stellar population model, in order to be consistent with the \citet{Goudfrooij14} interpretation of the intermediate age clusters. 
This, however, is in conflict with the interpretation by \citet{Milone15}}. The filled (red) triangle represents the results of \citet{BrandtHuang15a} for the Praesepe open cluster. These authors interpret the spread as being due to rotation, but show that if non-rotating isochrones are used, an age spread of $\sim200$~Myr is inferred.  

Finally, we show the results from \citet{Niederhofer15a} as upside-down (purple) triangles. While those authors did not interpret the derived age spreads as real, as they were largely interested in testing if such clusters had age spreads of 100s of Myr within them, they did find that the measured spreads were larger than expected due to photometric errors. 
We took their results and de-convolved the total dispersion from the expected spread due to photometric errors. Then we turned the resulting $1\sigma$\ dispersions to FWHM to more directly compare with the other works.
\citet{BastianSilva13} also studied NGC~1856 with the same method as \citet{Niederhofer15a}, namely investigating the SFH using the MSTO and post-MS features. These authors found a dispersion of 35~Myr, corresponding to a FWHM of $\sim85$~Myr. This is significantly shorter than that found by \citet{Milone15} and \citet{Correnti15}, with the apparent difference being due to the inclusion of the post-MS features, that show less variation than the MSTO.

As can be seen, the observational data closely follow the prediction from the rotating isochrones. Hence, the data add quantitative confirmation for a key prediction of the rotating isochrones, that the inferred age spread is proportional to the age of the stellar system.  This prediction can be further tested with high precision photometry of other young massive clusters in the LMC/SMC.  For example, the slightly younger ($\sim180$~Myr) but similar mass ($\sim10^5$~\msun) cluster, NGC~1866 \citep{BastianSilva13} should also show an eMSTO, but the extent of which should be less than NGC~1856 when translated into 'age space'.  Another massive ($>10^5$~\msun) young ($\sim100$~Myr) cluster that should show an eMSTO is NGC~1850, with the expected MSTO spread (when interpreted as an age spread) of $30-50$~Myr.

It is worthwhile noting that none of the clusters in this sample, or the larger sample of integrated spectroscopy examined by \citet{Bastian13}, show any evidence of ongoing star-formation.  Hence, if the spreads were real, all star-formation must have stopped recently, with no clusters caught in the act of forming new stars.

\section{Conclusions}
\label{sec:conclusions}

In this work we used the \textsc{Syclist} stellar evolution models to investigate the effect of stellar rotation on CMDs of star clusters at various ages. 
Comparing isochrones of non-rotating and rotating stars at different ages between 50 Myr and 1 Gyr 
we find that rotating stars, which have longer MS lifetimes due to enhanced internal mixing, can be mis-interpreted as younger stars. At the location of the MSTO, rotating isochrones closely follow younger non-rotating ones, causing a spread in the MSTO. This difference is proportional to the age of the isochrone in the considered age interval. In the light of the age spread interpretation, this would result in an increasing age spread with increasing age of a cluster. We compared literature measurements of apparent age spreads in young and intermediate age clusters with the predictions from the \textsc{Syclist} models. We found a strong correlation between the age of the clusters and the reported spread in ages, in close agreement with model expectations. 

We also investigated the effect that rotation has on the post-MS features of clusters and found that rotation produces evolved stellar tracks that are more compact in the CMD than the ones produced by the corresponding age spreads, although the MSTO region is similar in both cases, in agreement with the results inferred from young and intermediate age clusters \citep{BastianSilva13, Niederhofer15b}.

The model presented in this work that relates the eMSTO feature to rotation makes clear predictions that can observationally be tested.
If rotation is causing the spread in the MSTO, also two young clusters NGC~1850 and NGC~1866 should show an extended MSTO, as well. However, this extent should be smaller than the one found in NGC~1856, and of the order of $20-100$~Myr, if interpreted as an age spread. Moreover, our model predicts different post-MS morphologies in the age spread and rotation interpretation, making the evolved stellar part of the CMD a valuable tool to discriminate between the two scenarios.

\section*{Acknowledgments}
This research was supported by the DFG cluster of excellence "Origin and Structure of the Universe". \\
C.G. acknowledges support from the European Research Council under the European Union's Seventh Framework Programme (FP/2007-2013) / ERC Grant Agreement n. 306901. \\
NB is partially funded by a Royal Society University Research Fellowship. \\
We thank Paul Goudfrooij for useful comments. \\
We are grateful to the anonymous referee for useful comments and suggestions that helped to improve the manuscript.

\bsp
\label{lastpage}
\end{document}